\documentclass[aps,prl,twocolumn,superscriptaddress]{revtex4-1}
\usepackage{dcolumn}
\usepackage{bm}
\usepackage[T1]{fontenc}
\usepackage{braket}
\usepackage{amsmath}
\usepackage{amssymb}
\usepackage{nccmath}
\usepackage{mathrsfs}
\usepackage{pgf}   
\usepackage{graphicx}

\usepackage{xcolor}
\usepackage{lipsum}
\usepackage{array}
\newcolumntype{P}[1]{>{\centering\arraybackslash}p{#1}}
\let\oldsim\sim 
\renewcommand{\sim}{{\oldsim}}

\title{Control of reactive collisions by quantum interference}

\begin{document}
	\title{Control of reactive collisions by quantum interference}

	\author{Hyungmok Son}
    \email[Corresponding author \\ Email address: ]{hson@g.harvard.edu}
    \affiliation{MIT-Harvard Center for Ultracold Atoms, Research Laboratory of Electronics, 
    Department of Physics, Massachusetts Institute of Technology, Cambridge, Massachusetts 02139, USA}
    \affiliation{Department of Physics, Harvard University, Cambridge, Massachusetts 02138, USA}

    \author{Juliana J. Park}
    \affiliation{MIT-Harvard Center for Ultracold Atoms, Research Laboratory of Electronics, 
    Department of Physics, Massachusetts Institute of Technology, Cambridge, Massachusetts 02139, USA}
    
    \author{Yu-Kun Lu}
    \affiliation{MIT-Harvard Center for Ultracold Atoms, Research Laboratory of Electronics, 
    Department of Physics, Massachusetts Institute of Technology, Cambridge, Massachusetts 02139, USA}
    
    \author{Alan O. Jamison}
    \affiliation{Institute for Quantum Computing and Department of Physics $\&$ Astronomy, University of Waterloo, Waterloo, Ontario N2L 3G1, Canada}
	
    \author{Tijs Karman}
    \affiliation{Institute for Molecules and Materials, Radboud University, Heijendaalseweg 135, 6525 AJ Nijmegen, Netherlands}
	
    \author{Wolfgang Ketterle}
    \affiliation{MIT-Harvard Center for Ultracold Atoms, Research Laboratory of Electronics, 
    Department of Physics, Massachusetts Institute of Technology, Cambridge, Massachusetts 02139, USA}
        
	\begin{abstract}
    In this study, we achieved magnetic control of reactive scattering in an ultracold mixture of $^{23}$Na atoms and $^{23}$Na$^{6}$Li molecules. In most molecular collisions, particles react or are lost near short range with unity probability, leading to the so-called universal rate. By contrast, the Na{+}NaLi system was shown to have only $\sim4\%$ loss probability in a fully spin-polarized state. By controlling the phase of the scattering wave function via a Feshbach resonance, we modified the loss rate by more than a factor of $100$, from far below to far above the universal limit. The results are explained in analogy with an optical Fabry-Perot resonator by interference of reflections at short and long range. Our work demonstrates quantum control of chemistry by magnetic fields with the full dynamic range predicted by our models.
	\end{abstract}
	\textbf{\maketitle}	

\section*{Introduction}
\begin{figure}
    \centering
    \includegraphics[keepaspectratio, width = 86mm]{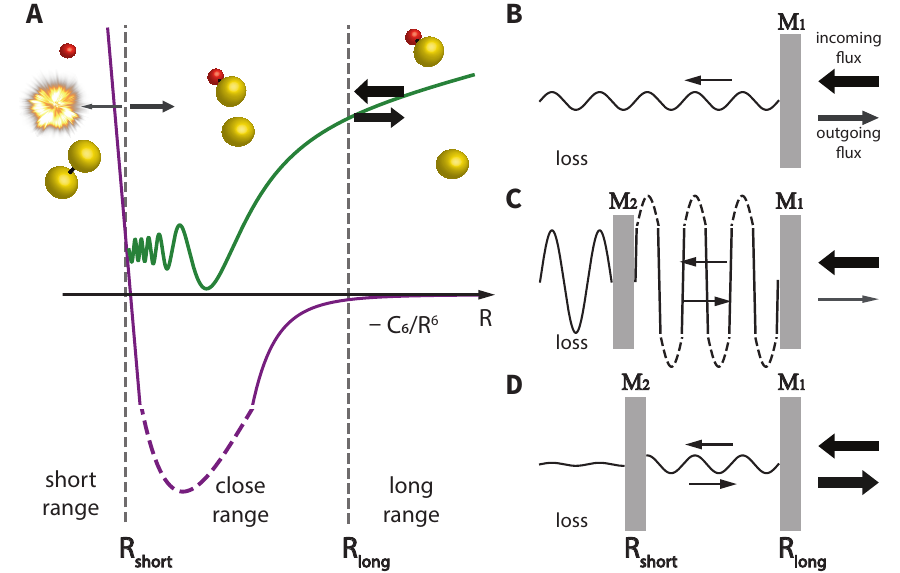}
    \caption{{\bf Fabry-Perot interferometer model for reactive collisions.} ($\textbf{A}$) Collisions between an atom (yellow sphere) and a molecule (yellow and red sphere together) occur in a potential that is the attractive vdW potential $-C_6/R^6$ at long range [$R{>}R_{\text{long}}$ (where $R$ is the interparticle distance)] and a strongly repulsive potential at short range ($R{<}R_{\text{short}}$). The scattering dynamics (represented by the wave function in green) can be fully described by quantum reflection off of the vdW potential at $R{\sim}R_{\text{long}}$ and reflection and transmission at $R{\sim}R_{\text{short}}$ \cite{PJ,PJ2}. Loss at close and short range is caused by the coupling of the incoming channel (in our case, a chemically stable quartet state) to a lossy channel (here, a reactive doublet state) and described by a reflectivity of ${<}100\%$ of the inner mirror. For instance, in the doublet state, a singlet Na$_2$ molecule (two yellow spheres) can be formed. This situation is fully analogous to an optical Fabry-Perot interferometer with two partially reflective mirrors (M$_1$ and M$_2$). Reactive loss is proportional to the flux transmitted through both mirrors. ($\textbf{B}$) Transmission through mirror M$_1$ only represents the universal loss. Depending on constructive and destructive interference between multiple reflections, the loss can be highly enhanced [on resonance ($\textbf{C}$)] or suppressed ($\textbf{D}$) relative to the universal loss. (C) and (D) are depicted for a reflectivity $|r_2|^2 {\sim} 0.89$, with 1000-fold loss enhancement between these panels.}
    \label{fig:cartoon}
\end{figure}

Advances in cooling atoms and molecules have opened up the field of quantum scattering resonances \cite{firstfeshbach,chengchin_feshbach,res_narevicius,no_he_res} and ultracold chemistry \cite{ColdMolReview,krbrecent_yliu}. At micro- and nanokelvin temperatures, collisions occur only in the lowest partial wave, and the collisional physics can be reduced to a few well-defined parameters, which, in many systems, are the $s$-wave scattering length and a two-body loss-rate coefficient. Collisions involving molecules are often much more complex, owing to the strong anisotropic interaction at short range and multiple decay channels including reactions \cite{statres_bohn,nak_photo,rbcs_photo,yliu_complexlifetime}. One goal of current research is to identify systems that can still be understood with relatively simple models, including the model of universal rate coefficients \cite{PJ}, as well as single-channel and two-channel models. Such systems are most likely to enable researchers to achieve controllable quantum chemistry in which the outcome of reactions is steered by external electromagnetic fields \cite{chem_external_krems,coldchem_softley}.

Collisions in molecular systems can be described by the reflection of the wave function in two regions. At long range, the attractive van der Waals (vdW) potential acts as a highly reflective mirror, owing to quantum reflection in low-temperature scattering (Fig.$~1$). When the colliding particles are in close proximity, they can again be reflected by the repulsive short-range potential, or they can get lost because of reactions and/or inelastic transfer to other states. These losses can be represented by transmission through the short-range mirror (Fig.$~1$). In the universal limit, the transmission is $100\%$, and the entire incoming flux is lost.

Most ultracold molecular collisions studied thus far have loss-rate coefficients at or close to the universal value \cite{krb_kkni,ZwierleinNaK,narb, rbcs_below_univ,rbcs_hanns,caf,rb2,li2,rbcscs_higher_similar,cs2cs,cs2cs_2,liscscs}. However, when partial reflection occurs at short range, the resulting loss rate can be higher or lower than the universal value, depending on the interference created by multiple reflection pathways. This fact is analogous to an optical Fabry-Perot interferometer. This optical analog fully captures the results of a single-channel description of reactive molecular collisions \cite{PJ,PJ2}. We extended the single-channel model by adding a Feshbach resonance as a lossless phase shifter that can tune between constructive and destructive interferences.

Our experimental system, which consists of collisions of triplet ro-vibrational ground-state NaLi with Na near $978\; \text{G}$, is fully described by this Fabry-Perot model. We saw loss rates that exceeded the universal rates by a factor of $\sim5$, tunable via a Feshbach resonance over a range of more than two orders of magnitude. We have also characterized a weaker loss resonance, where the phase shifter was ``lossy''—i.e., the closed channel of the Feshbach resonance had a short lifetime and dominated the loss, almost completely spoiling the quality factor of the Fabry-Perot resonator. This resonance has to be described by a two-channel model with the lifetime of the bound state as an additional parameter \cite{JH}. Our experiment has established NaLi${+}$Na as a distinctive system that realizes the full dynamic range of recent models developed to describe reactive collisions involving ultracold molecules \cite{PJ,PJ2}. Furthermore, to the best of our knowledge, this system is only the second example for which Feshbach resonances between ultracold molecules and atoms have been found \cite{nak_k_recent,nakk_binding}.

\section*{Experimental protocol}

A mixture of ${\sim}3{\times} 10^{5}$ sodium atoms and ${\sim} 3{\times}10^4$ NaLi molecules in the triplet ro-vibrational ground state was produced in a $1596$-nm one-dimensional (1D) optical lattice created by retroreflecting the trapping beam. The sample was confined as an array of $\sim1000$ pancake-shaped clouds. The atoms and molecules were both in the upper stretched hyperfine states, where all electron and nuclear spins were aligned along the bias field \cite{NaLiSympCool,NaLiGround}. This spin-polarized mixture was in a chemically stable quartet state. The sample was prepared at a temperature $T_{\text{NaLi}} {\approx} T_{\text{Na}} \sim T {=} 1.55\; \mu \text{K}$, well within the regime for threshold behavior of collisions, which required the temperature to be much less than the characteristic temperature determined by the vdW potential, $T_{\text{vdW}} {=} \hbar^2/2\mu k_B r^2_6$ \cite{vdwtemp} with the vdW length $r_6{=}(2\mu C_6/\hbar^2)^{1/4}$, where $\hbar$ is Planck’s constant divided by $2\pi$, $\mu$ is the reduced mass of NaLi and Na, $k_B$ is the Boltzmann constant, and $C_6$ is the vdW constant for the atom-molecule potential. With $C_6 {=} 4026$ in atomic units \cite{tscherbul_NaLiNa},
$T_{\text{vdW}} {\approx} 500\; \mu\text{K}$.

The atom-molecule mixture was initially prepared near the Na-Li Feshbach resonance at $745\ \text{G}$, and then the bias field was ramped to the target value in $15\; \text{ms}$. We determined collisional lifetimes of the atom-molecule mixture by holding the sample for a variable time at the target magnetic field, after which the field was ramped back to $745\ \text{G}$ where the remaining molecules were dissociated and detected. The number of dissociated Li or Na atoms in the hyperfine ground state was measured by resonant absorption imaging \cite{NaLiSympCool}. The hyperfine state of Na atoms from dissociation differed from that of Na atoms in the initial atom-molecule mixture.

Decay curves for the molecules are compared in the inset of Fig.$~2$, near and far away from the strong atom-molecule Feshbach resonance studied in this work. Our observable $\gamma(B)$ (where $\gamma$ is the loss rate and B is the magnetic field) is the difference of initial loss rates of NaLi molecules with and without Na atoms. It is obtained by fitting the whole decay curve using the standard differential equations for two-body decay [see supplementary materials (SM)]. The loss in the absence of Na atoms is caused by $p$-wave reactive collisions between molecules. The measured molecular two-body loss-rate coefficient $\beta {=} 2.6(7){\times} 10^{-12}\; (\text{cm}^{3}/\text{s}) (T_{\text{NaLi}}/\mu \text{K}) $ near $980\; \text{G}$ is within a factor of 2 of the prediction from the universal loss model (see SM). Single-particle loss due to the vacuum limited lifetime of ${>}20\; \text{s}$ was negligible.

We avoided the need for absolute sodium density measurements by comparing the measured decay rate to the decay rate of the mixture in a non-stretched spin state. Because this mixture collides on a highly reactive doublet potential, the decay was reliably predicted to occur with the $s$-wave universal rate coefficient, which is well known for our system \cite{tscherbul_NaLiNa}. By comparing datasets taken at different times (see SM), we estimated the uncertainty of the density calibration to be $\sim40\%$.

We also measured the loss rate of the mixture in the non-stretched spin state by using the measured particle number and temperature, as well as a model for the anharmonic trapping potential. With the measurement, we have confirmed within $30\%$ uncertainty that the rate is indeed the universal rate (see SM). Because we regard the theoretical prediction to be highly reliable, we did not use the experimental density calibration in the analysis reported here.

\begin{figure}
	\centering
	\includegraphics[width = 86 mm, keepaspectratio]{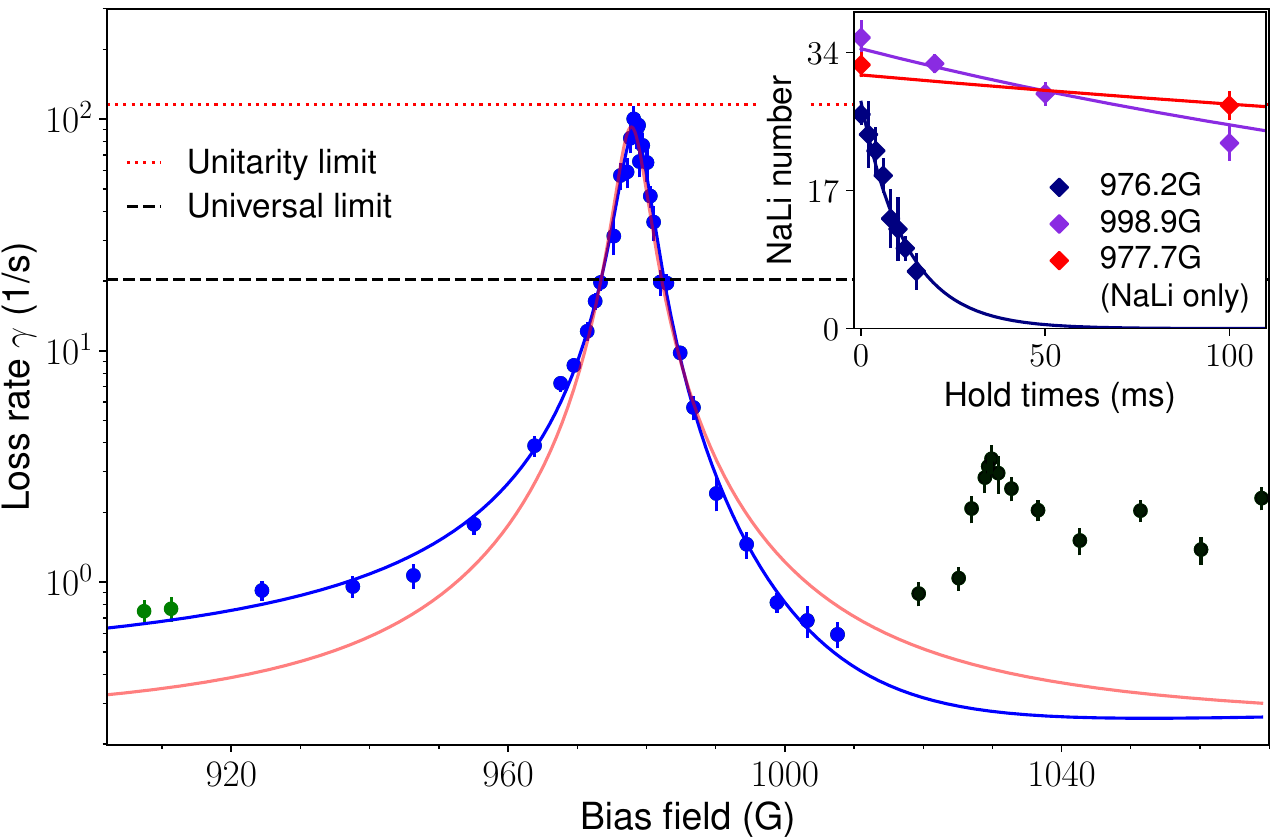}
	\caption{{\bf Observation of Feshbach resonances in Na${+}$NaLi collisions.} Observed decay rates are shown as a function of bias field, with $100$ Na atoms per pancake-shaped cloud at temperatures of $T_{\text{Na}} {=} 1.60\; \mu\text{K}$, $T_{\text{NaLi}} {=} 1.68\; \mu\text{K}$, corresponding to an overlap density of $1.1\times10^{11}\; \text{cm}^{-3}$. Data points taken with different sodium numbers and temperatures were scaled to the same overlap density (see SM). The blue line is a fit of the line shape to the Fabry-Perot contrast function $C$; the red line is a Lorentzian fit. Both fits use only the data points represented by blue circles. Green circles were excluded owing to another resonance near $880$ G. Black circles show an additional resonance near $1030$ G. The red dotted line is the unitarity limit in two dimensions for our experimental conditions. The black dashed line is the universal loss rate. Data points were acquired with 6 to 11 different hold times at each bias field; four to eight measurements at a given hold time were averaged. Error bars indicate 1 SD. The inset shows decay curves of molecules: The dark blue (or red) diamonds are near the strong resonance at $978$ G with (or without) Na atoms, and the purple diamonds are off resonance near $1000$ G with atoms. Relative to the dark blue diamonds, the overlap density of the purple data is larger by a factor of $2.2$. The red line in the inset is a fit for the two-body molecular loss. The dark blue and purple lines are obtained by fitting the standard differential equations of two-body decay processes, including collisions with atoms (see SM).}
	\label{fig:figure2}
\end{figure}

\section*{Fabry-Perot interferometer model}
Reactive scattering between molecules and atoms can be matched to the simple picture of an optical Fabry-Perot interferometer with two reflectors, M$_1$ and M$_2$ (Fig.~1). Mirror M$_1$ represents quantum reflection by the long-range vdW potential, and M$_2$ represents reflection near short range. Inelastic and reactive losses, which occur at close or short range (Fig.~1), are represented in the Fabry-Perot picture by transmission through the inner reflector M2, followed by absorption. For an incoming flux $I$, the total transmission $T_{tot}$ through both reflectors is given by
\begin{equation}
    T_{tot} = (I \cdot |t_1|^2) \bigg(\frac{1-|r_2|^2}{|1 - r_2r_1e^{-i\phi}|^2}\bigg) \equiv (I \cdot |t_1|^2) C
    \label{eq:fabryperot}
\end{equation}
$r_i$ and $t_i$  are the amplitude reflection and transmission coefficients for mirror $M_i$, and $\phi$ is the round-trip phase, which, in the Fabry-Perot model, can be tuned by the distance between mirrors or the refractive index of the medium. The term $I \cdot |t_1|^2$ is the transmitted flux in the absence of the inner mirror (i.e., $r_2{=}0$) and, for collisions, represents the universal loss. The factor C represents the effect of interference. For later convenience, we characterize the inner reflection by a parameter $0 \leq y \leq1$: $r_2 =(1-y)/(1+y)$, $t_2= 2\sqrt{y}/(1+y)$, 
which is 1 for complete transmission and 0 for complete reflection. With $r_1 \sim 1$ (quantum reflection approaches unity at low energies), we obtain $C(y, \phi)= 2 y/(1 - \cos{\phi} + y^2 (1 +\cos{\phi}))$. Constructive interference at $\phi = 0$ leads to an enhancement $C=1/y$, and destructive interference at $\phi = \pi$ leads to a minimum transmission with $C=y$. In the limit of small $y \ll 1$ relevant for our experimental results, the transmission probability for the inner mirror is $\sim4y$.

In the case of cold collisions, scattering rates are periodic when the close-range potential is modified and new bound states are added to the interparticle potential. Each new bound state results in a resonance and ``tunes'' the Fabry-Perot interferometer over one full spectral range with the scattering length $a$ varying by $\pm\infty$. In accordance with \cite{PJ}, we defined the normalized scattering length $s=a/\bar{a}$, where $\bar{a} = 0.47799\cdot r_{6}$  the mean scattering length \cite{mean_scatt_length}. If we substitute $\cos{\phi}=1-2/(1+(1-s)^2)$\cite{note0}, we obtain $C(y,s)=y (1+(1-s)^2)/(1+y^2(1-s)^2)$. This expression exactly reproduces the results of the quantum-defect model used in \cite{PJ} for the imaginary part of the scattering length, $\beta = \bar{a} C(y, s)$ which is proportional to the zero-temperature loss-rate coefficient. Previous studies \cite{PJ,PJ2,cong_qdt} already pointed out that their results can be interpreted as an interference effect of multiple reflections between short and long range. The relation between the phase shift $\phi$ and the parameter $s$ exactly reflects how a short-range phase shift modifies the scattering length \cite{PJ2}.

We extended this single-channel model (where $s$ is the normalized background $s$-wave scattering length without loss, $y{=}0$) by adding a Feshbach resonance as a lossless phase shifter for the Fabry-Perot phase
\begin{equation}    
    s(B) = q\bigg(1 - \cfrac{\Delta}{B-B_{\text{res}}}\bigg)
\end{equation}    

The resonance is at a magnetic field $B{=}B_{\text{res}}$, $q$ characterizes the background scattering phase far away from the Feshbach resonance, and $\Delta$ is the width of the resonance. Tuning the magnetic field across the resonance takes the Fabry-Perot interferometer across a full spectral range and provides tunable interference at fixed low temperature. For finite collision energies, interference has been observed also as a function of collision energy \cite{xie_hdd}.

\section*{Results and analysis}
The measured loss rates for Na${+}$NaLi collisions as a function of magnetic field are shown in Fig.~2. These data, which reveal a resonant enhancement of the loss by more than two orders of magnitude, represent the main result of this paper. Because the ratio of maximum and minimum loss is $y^2$ in the Fabry-Perot model, this result immediately suggested that $y$ has to be $<0.1$. We could fit the asymmetric line shape of the resonance well to the function $C(y, s(B))$ with an overall normalization factor and obtain $y{=}0.05$ and $q{=}1.61$. However, the observed peak losses were close to the unitarity limit, which provides an upper limit for elastic and inelastic scattering rates. When the scattering length exceeds the de Broglie wavelength $\lambdabar = 1/k$ (where $k$ is the relative wave number), the elastic cross section in 3D saturates at $4\pi\lambdabar^2$, whereas the inelastic rate coefficient peaks at $(h/2\mu)\lambdabar$ (where $h$ is Planck’s constant). The observed peak loss rate was close to the unitarity limit, so we had to consider the role of nonzero momentum.

Combining the threshold quantum-defect model for the complex scattering length with a finite momentum S-matrix formulation for the scattering rates, Idziaszek and Julienne \cite{PJ} obtained the complex scattering length
\begin{equation}    
    \tilde{a} = \bar{a} \bigg(s + y \cfrac{1 + (1-s)^2}{i + y(1-s)}\bigg) \equiv \alpha -i\beta
\end{equation}  
where $\alpha$ is the real and $\beta= \bar{a} C(y,s(B))$ is the imaginary part of $\bar{a}$. The loss-rate coefficient $K$ is given by
\begin{equation}
    K = f(k)\cfrac{2h}{\mu}\beta
	\label{eq:lossconst}
\end{equation}      
The function $f(k) = (1 + k^2|\tilde{a}|^2 + 2k\beta)^{-1} $
establishes the unitarity limit for the scattering rates. So far, we have discussed inelastic scattering in three dimensions. However, because our atomic and molecular clouds had the shape of thin pancakes, we were in a 2D regime. Because the vdW length was much smaller than the thickness of the pancakes, the collisions were microscopically 3D and described by the 3D complex scattering length, but additionally, one had to use 2D scattering functions, leading to two effects. First, there is a logarithmic correction of the scattering length. For harmonic axial confinement with frequency $\omega_{ax}$, the correction factor is $l {=}|(1 {+} (\tilde{a}/\sqrt{\pi}l_o)\text{ln}(B\hbar\omega_{ax}/\pi k_B T))|^{-2}$ where $B{\approx}0.915$, and $l_o{=}\sqrt{\hbar/\mu\omega_{ax}}$ is the associated oscillator length \cite{dima_2d,ijj_2d}. Because Na and NaLi have different axial confinement frequencies $\omega_{i}$ and masses $m_i$, we used $\omega_{\text{ax}} = \mu(\omega_{\text{Na}}/m_{\text{Na}}+\omega_{\text{NaLi}}/m_{\text{NaLi}})$\cite{Note1}. The correction factor $l$ is large only at extremely low temperatures and near confinement-induced resonances \cite{dima_2d,ijj_2d}. In our case, it provided a small shift of the peak loss by ${\sim} 0.3\; \text{G}$. The second modification resulting from the 2D nature of the confinement is in the saturation factor, $f(k)$: In three dimensions, $k$ is obtained from the thermal energy, whereas in two dimensions it is obtained from $2\pi$ times the relative kinetic energy of the zero-point motion, $k{=}\sqrt{\pi}/l_o$. The factor of $2\pi$ is a reminder that 2D dynamics cannot be fully captured by adding the zero-point energy to the thermal energy.

Our density calibration used the loss rate for collisions in a Na${+}$ NaLi mixture in a nonstretched spin state (see the “Experimental protocol” section). The loss rate is expressed by the imaginary part of the scattering length, $\beta$, and for the universal rate, $\beta {=} \bar{a}$.

For the ratio of the observed loss rate to the loss rate measured for the nonstretched state, we obtain
\begin{equation}
    \begin{split}
     r(B)&=f(k) l (\beta/ \bar{a}) \\
         &=\frac{|1 {+} (\tilde{a}/\sqrt{\pi}l_o)\text{ln}(B\hbar^2/\pi\mu l_o^2k_B T)|^{-2}}{1 + (\sqrt{\pi}/l_{o})^2l|\tilde{a}|^2+2(\sqrt{\pi}/l_o)l\beta}\bigg(\cfrac{\beta}{\bar{a}}\bigg)
    \end{split}
\end{equation}
The sodium density and all other factors are common mode and are canceled by taking the ratio. For the calibration measurement, $f(k) = 1$ and $l = 1$, owing to the smallness of the scattering length.

\begin{figure}
	\centering
	\includegraphics[width = 86mm, keepaspectratio]{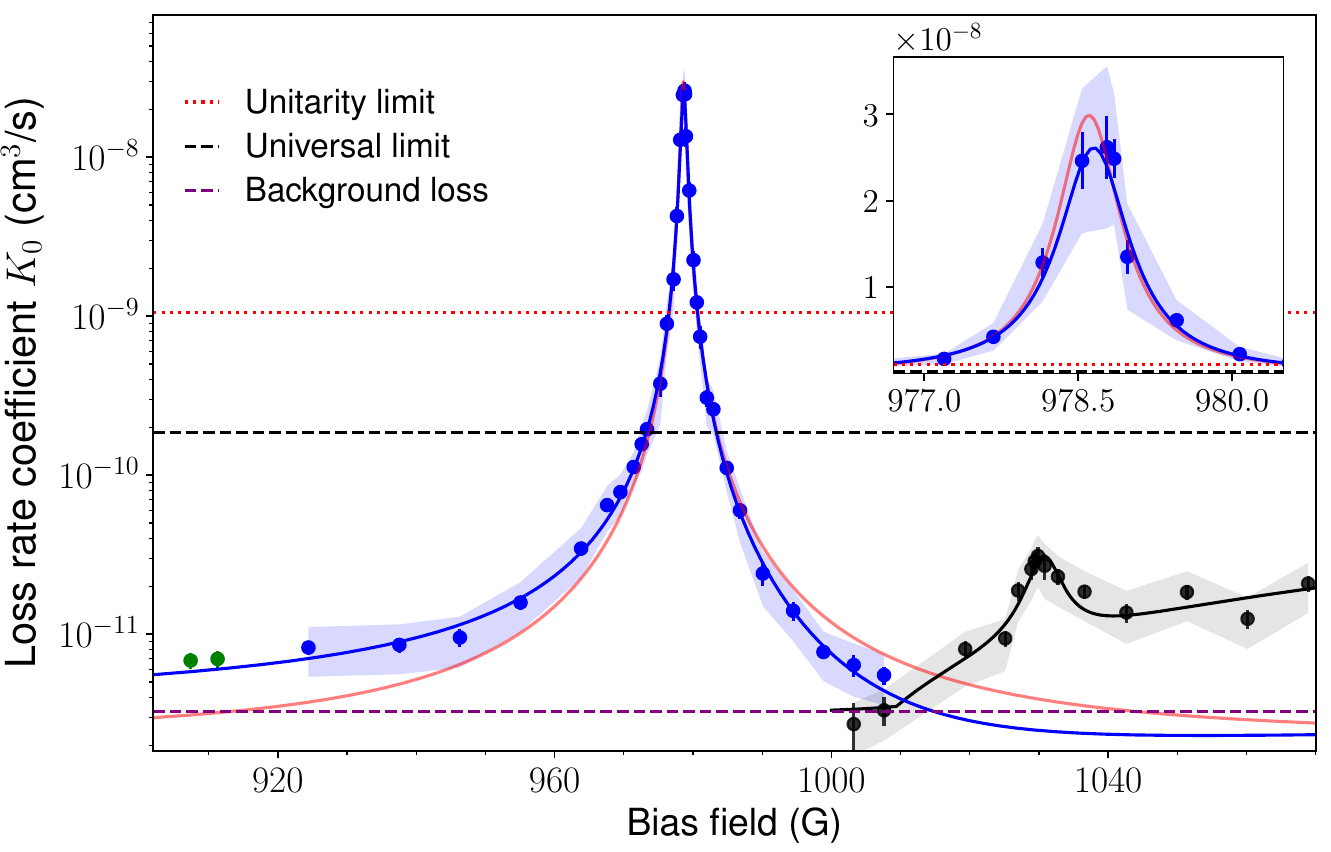}
	\caption{{\bf Zero-temperature loss rate coefficients $K_{0}$ for Na${+}$NaLi collisions.} $K_{0}$ is the imaginary part of the scattering length times $2h/\mu$. Experimental data points were corrected for nonzero momentum effects, $f(k)l$ (see the “Results and analysis” section). The blue line is the best fit based on the single-channel model; the red line is a symmetric Lorentzian fit. For both fits, only blue data points were included. The black line is a fit of the weak resonance using a two-channel model. The two black data points near $1005$ G were obtained from the blue points by subtracting the contribution of the wings of the strong resonance. The red dotted line is the unitarity limit in two dimensions, $(h/2\mu)\sqrt{\pi}/l_o$ The black dashed line is the universal limit. The purple dashed line shows the background (open-channel) loss. The shaded area represents the total uncertainty, which is the quadrature sum of the standard deviation and the systematic uncertainty in the density calibration. The inset shows a magnified view of the central part of the figure on a linear scale.}
	\label{fig:figure3}
\end{figure}
We fit the loss-rate ratio $r(B)$ using four parameters: $B_{\text{res}}, \Delta, q,$ and $y$. Because the calibration measurements had uncertainties, we included a fifth fitting parameter in the form of an overall normalization factor, $\mathcal{N}$. We used an accurate theoretical value of $\bar{a}$ calculated for triplet ground-state NaLi${+}$Na: $\bar{a} = 56.1 a_{0}$ with $\lesssim 1.5 \%$ uncertainty \cite{tscherbul_NaLiNa}. Figure~3 compares the experimental results with the fits. In the figure, we have multiplied $r(B)$ by the constant $2h\bar{a}/\mu$ and divided by the momentum-dependent 2D corrections $f(k)l$ calculated with the parameters of the best fit. In this way, we obtained the zero-temperature 3D loss-rate coefficient, $K_{0}(B)=(2h/\mu) \beta$, which is a microscopic property of the two-body system Na${+}$NaLi.

The best fit with the single-channel model yielded $B_{\text{res}}{=}978.6(1)\; \text{G}$, $\Delta {=} 28 (2)\; \text{G}$, $y {=} 0.0094(47)$, $q {=} 1.60(7)$ with $\chi^2_{\text{red}} {=} 1.23$  [degrees of freedom (dof) = $27$]. The normalization factor $\mathcal{N} {=} 1.32(45)$ was compatible with 1 and therefore was consistent with our calibration method.

Figure~3 also shows a weaker resonance near $1030\; \text{G}$. This resonance could not be explained by the single-channel model (i.e., with a lossless phase shifter), which predicts that the maximum loss is larger than the universal limit. We therefore extended our model by considering a finite lifetime of the bound state coupled by the Feshbach resonance in the form of a linewidth $\Gamma_{b}$. This extension implied that the phase shifter of the Fabry-Perot resonator was now lossy and prevented the large resonant buildup of wave function inside the interferometer. In this two-channel model, we describe a lossy phase-shifter by
\begin{equation}    
    s_{b}(B) = q\bigg(1 - \cfrac{\Delta}{B-B_{\text{res}}-i\Gamma_{b}/2}\bigg)
	\label{eq:phaseshifter_loss}
\end{equation} 
With Eq.~(\ref{eq:phaseshifter_loss}), we can express $\beta$ in the Breit-Wigner form as in \cite{JH} (see SM). We assumed that $y$ and $q$ were the same as for the strong resonance because they characterized the same incoming channel, and we used the same normalization factor. We also included a slope and an offset as extra fit parameters to account for other loss channels not covered by our model. We obtained $B_{\text{res}}{=}1030.8(7)\; \text{G}$, $\Delta {=}0.21(4)\; \text{G}$, $\Gamma_{b}{=}5.38(4)\; \text{G}$ with $\chi^2_{\text{red}}{=}2.4$ (dof = $10$). The $\Delta$ parameter showed that the resonance at $1030\; \text{G}$ was two orders of magnitude weaker than the one at $978\; \text{G}$. From the linewidth, we inferred the bound-state lifetime $\tau_{b} = (\delta \mu \Gamma_{b})^{-1} \approx 60\; \text{ns}$, where the relative magnetic moment between the entrance channel and the closed-channel bound state was $\delta\mu {=} 2\mu_B$ (where $\mu_B$ is the Bohr magneton) assuming a single spin-flip. In our study, the lifetime of a collision complex was obtained from a spectroscopic linewidth, whereas in all previous work on collisions of ultracold molecules, such lifetimes were obtained from a direct time-domain measurement \cite{yliu_complexlifetime,rbcs_photo}. The short lifetime of the bound state suggests that the closed channel has a highly reactive doublet character. Some contribution to $\Gamma_{b}$ and $y$ could also come from the $1596$-nm trapping light, which can excite collision complexes leading to loss, as observed in other molecular systems \cite{nak_photo,rbcs_photo,yliu_complexlifetime}. However, given the small value of $y$, we expect this effect to be small.

\begin{figure}
	\centering
	\includegraphics[width = 86mm, keepaspectratio]{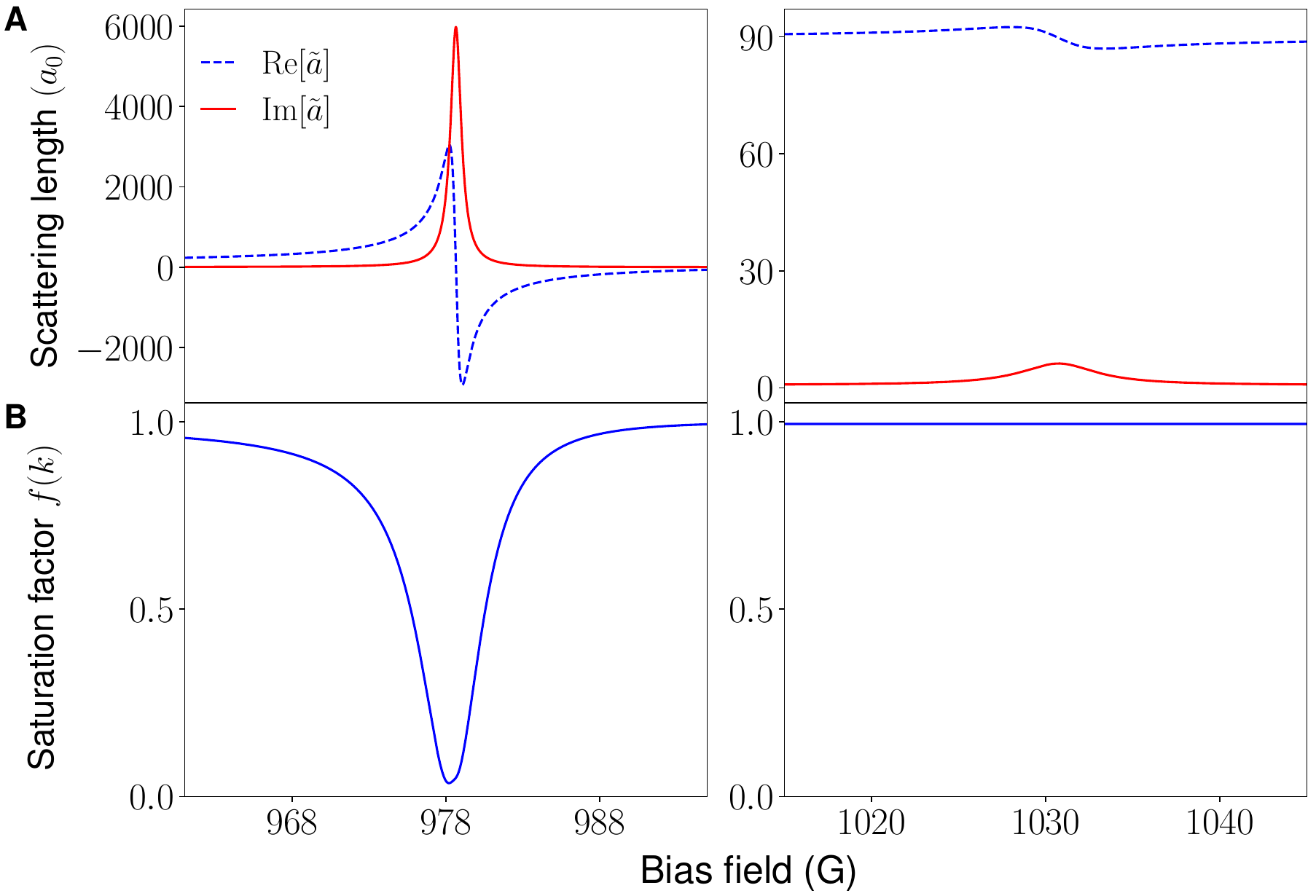}
	\caption{{\bf Complex scattering length and saturation factor calculated from the best-fit results.} (\textbf{A}) Real (Re) and imaginary (Im) parts of the scattering lengths, $\tilde{a}$ near the strong resonance calculated based on the single-channel model (left plot), and near the weak resonance based on the two-channel model (right plot). (\textbf{B}) Saturation factor $f(k)$ with $k{=}\sqrt{\pi}/l_o$ where $l_o$ is the oscillator length for the effective axial confinement frequency. The saturation was negligible for the weak resonance (right plot), whereas $f(k){\sim} 0.03$ at the strong resonance (left plot).}
	\label{fig:figure4}
\end{figure}

We could also fit the strong resonance to the two-channel model and found $\Gamma_{b} = 0 {\pm} 1\; \text{G}$, confirming that we can regard the strong resonance as a lossless phase shifter. The longer lifetime of the closed channel associated with the strong resonance suggests that it is only weakly coupled to reactive channels. The difference between the two Feshbach resonances is highlighted by examining the total inelastic width, $\Gamma_{\text{inel}}$, of the resonance (see SM)
\begin{equation}
\Gamma_{\text{inel}} = \Gamma_{\text{b}} + \cfrac{2yq\Delta}{1 + y^2(q-1)^2}
\end{equation}
where the first term is the natural linewidth of the bound state itself, and the second term represents the resonantly enhanced decay rate of the incoming channel. The width of the 978-G resonance was dominated by the open-channel losses at short range (i.e., $y \ne 0$), whereas the weak resonance was limited by the decay rate $\Gamma_{\text{b}}$. Equation (7) shows that the y parameter is more easily determined from a strong resonance. By contrast, the weaker resonance was insensitive to the short-range parameters $y$ and $q$ of the incoming channel. Figure~4 shows the real and imaginary parts of the scattering lengths for the two resonances and illustrates the power of the simple model: Analysis of the inelastic scattering provides a full description of all $s$-wave scattering properties, including elastic scattering and momentum dependence. We can also calculate the good-to-bad collision ratio, $k(\alpha^2+\beta^2)/\beta$, and find that it is maximized away from the resonance (see SM).

Figure~3 shows that the zero-momentum loss rate could be tuned over four orders of magnitude and exceeded the universal limit by a factor of $100$, which was reduced by the unitarity limit to a factor of $5$ (Fig.~2) [see also \cite{Note2}].

The quality factor of the Fabry-Perot resonator becomes smaller for nonzero momentum, owing to the lower long-range quantum reflectivity, which has a threshold law of $|r_1| \approx 1{-}2 \bar{a} k$. This relation yields $|r_1| {\sim} 0.93$ using the total (i.e., thermal and zero-point) momentum for $k$. The resonant enhancement inside the Fabry-Perot resonator is reduced when the transmission of the outer mirror (M$_1$) is comparable to that of the inner mirror (M$_2$)—i.e., when $\bar{a} k \sim y$. At this point, the unitarity saturation takes effect and reduces the loss rate from its zero-temperature value shown in Fig.~3.

\section*{Discussion} 
In this study, we have demonstrated the substantial suppression and enhancement of reactive collisions relative to the universal limit, which is possible only if $y {\ll} 1$, and we have achieved control of chemical reactions via external magnetic fields. An asymmetric line shape can lead to a suppression of inelastic losses below the background loss \cite{JH_lossreduction}. This suppression was not realized for the results shown in Fig.~3, owing to the neighboring weaker Feshbach resonance.

Our analysis highlights the conditions necessary to observe such a high dynamic range tunability of reactive collisions. The possible contrast is given by $1/y^2$ but is only realized if the Feshbach resonance is sufficiently strong and coupled to a sufficiently long-lived state: $q \Delta/\Gamma_{\text{b}} {>} 1/y$. This condition for the Feshbach resonance is more difficult to fulfill for smaller values of $y$, but the Na${+}$NaLi system satisfies this condition for the resonance at $978\; \text{G}$ and for several other resonances that we have observed but not yet fully analyzed.

The models for reactive collisions presented here may look rather specialized. However, our two-channel model captures the low-temperature limit of the most general resonance possible for which the complex scattering length is represented by a circle in the complex plane \cite{JH,JH_detail} (see SM).

Universal reaction rates are determined only by quantum reflection of the long-range potential and do not provide any information about the ``real chemistry'' at short range. Therefore, discovery and characterization of nonuniversal molecular systems are major goals of the field \cite{ZwierleinNaK,narb,rbcs_below_univ,caf_mwshielding,li2,rb2}. However, most of the cases studied exhibited only two- to fourfold deviation from the universal limit, and interpretation of these cases required an accurate density calibration that was not always performed. Some studies showed inelastic rates well below the universal limit, without any resonances \cite{caf_rb,bosonic_nakk,li2_tout}, which can provide only an upper bound for $y$ and leave $q$ undetermined. This work has demonstrated how short-range reflectivity makes it possible to access information about short-range interactions and collisional intermediate complexes. Our analysis showed that suppression of loss below the universal limit could occur for a wide range of parameters, but strong enhancement of loss beyond the universal limit requires fine tuning: an almost lossless Fabry-Perot interferometer tuned to resonance.

In this work, we have experimentally validated a method on the basis of external magnetic fields and quantum interference to realize quantum control of chemistry. Previous studies used microwaves \cite{nak_mw,caf_mwshielding} or electric fields \cite{krb_efield_2d} to control losses in molecular systems with strong long-range dipolar interactions by modifying the universal rate limit. In our study, we used magnetic fields and quantum interference, without the need for dipolar interactions, to achieve loss-rate coefficients that far exceed the universal limit. All of these methods control one specific decay channel. With the weak resonance, we have also demonstrated that magnetic field can switch between two different mechanisms of reactive scattering, occurring in the chemically stable incoming and the lossy closed channels, respectively.

\newpage

\begin{thebibliography}{10}
\bibitem{note0}
This expression is equivalent to $\text{tan}(\phi/2)=1/(1-s)$, which is also
  given in Ref.~\cite{PJ2}.
\end{thebibliography}

\bibliographystyle{apsrev4-1}

\newpage
\section*{Acknowledgements} 
We would like to thank Dmitry Petrov, Paul Julienne, Krzysztof Jachymski, and Timur Tscherbul for valuable discussions. {\bf Funding:} We acknowledge support from the NSF through the Center for Ultracold Atoms and Grant No. 1506369 and from the Air Force Office of Scientific Research (MURI, Grant No. FA9550-21-1-0069). Some of the analysis was performed by W.K. at the Aspen Center for Physics, which is supported by National Science Foundation grant PHY-1607611. H.S. and J.J.P. acknowledge additional support from the Samsung Scholarship. {\bf Author contributions:} H.S. and J.J.P. carried out the experimental work. All authors contributed to the development of models, data analysis, and writing the manuscript. {\bf Competing interests:} None declared. {\bf Data and materials availability:} All data needed to evaluate the conclusions in the paper are present in the paper or the Supplementary Materials. All data presented in this paper are deposited at Zenodo \cite{ourdata}.


\newpage
\clearpage
\section{Supplemental Materials}
\subsection{Loss rate measurement $\&$ density calibration}
\noindent In the main article, we provided the basic concept of how we acquired loss rates for Na${+}$NaLi collisions, by comparing the initial loss rates with and without sodium atoms. For better accuracy, we evaluated the whole decay curve of molecules taking the time-dependent sodium numbers and particle temperatures into account, by numerically solving the differential equations:
\begin{align}
    &\dot{N}_{\text{NaLi}} = -(K/V_{\text{ov}})N_{\text{Na}}N_{\text{NaLi}} - (\beta/V_{\text{eff}})N^2_{\text{NaLi}} \label{eq:diffeqn1}\tag{S1} \\
    &\dot{N}_{\text{Na}} = -(K/V_{\text{ov}})N_{\text{NaLi}}N_{\text{Na}}, \label{eq:diffeqn2}\tag{S2}
\end{align}
\noindent where $N_{i}$ represents the effective number of particles of type $i$ per pancake,
$V_{\text{ov}}$ is the volume of the regime where atoms overlap with the more tightly confined molecules, and $V_{\text{eff}}$ is the mean volume of molecules. $\beta$ is the two-body molecular loss rate coefficient and
 $K$ is the loss rate coefficient for the collisions of Na${+}$NaLi pairs.
 
 To determine rate coefficients from equations Eq.~(\ref{eq:diffeqn1}) and Eq.~(\ref{eq:diffeqn2}) requires accurate knowledge of the volumes $V_{\text{eff}}$ and $V_{\text{ov}}$ for a single pancake and knowledge of the effective particle number per pancake, which is the weighted average over the entire ensemble of pancakes. We will discuss this below.
 
 In our primary calibration, however, we avoided this complication by comparing the decay of the mixture in the stretched spin state around the Feshbach resonance to that in a non-stretched spin state where the rate coefficient $K$ is reliably predicted to be the $s$-wave universal loss rate coefficient. For this, we prepared sodium atoms in the lowest hyperfine state ($\ket{F, m_F} = \ket{1,1}$ at the low field) and molecules are in their upper spin stretched state. For these initial states, the Na${+}$NaLi system has a $50 \%$ doublet character. The doublet potential at short range is highly reactive due to exothermic nature and the collisional loss is predicted to occur at the $s$-wave universal rate $(\textit{33})$. In other words, we compared the decay of samples in the spin stretched and non-stretched states with the same initial densities, i.e. with the same effective particle numbers and the same volumes $V_{\text{eff}}$ and $V_{\text{ov}}$. The only difference was the rate coefficient $K$, and by comparing the two decay curves, we obtained an accurate value for the ratio of the two $K$ coefficients without an accurate model for the volumes and the ensemble of pancakes. Multiple calibration measurements with the non-stretched spin state mixture suggest a total uncertainty of the density calibration of ${\sim} 40 \%$. 

We have validated the assumption of the universal decay of the mixture in the non-stretched state by carrying out an independent absolute calibration. In this calibration, we obtained the effective particle numbers from the measured numbers of atoms and molecules and their distributions over pancakes, and the overlap volume using a full anharmonic model of the trapping potential and measured temperatures. For the accurate estimation, we had to sum the decay curves over the ensemble of pancakes, or alternatively, calculate the effective particle numbers per pancake and consider an effective decay curve using them. Since the observed axial profiles over pancakes of both Na and NaLi followed Gaussian fits with widths $\sigma_{\text{NaLi}}{=}450(60) \mu\text{m}$ and $\sigma_{\text{Na}}{=}630(60) \mu\text{m}$, we assumed a Gaussian distribution of the particle number per pancake.  The particle numbers in the central pancakes, $N_{\text{Na}}^0$ and $N_{\text{NaLi}}^0$, are related to the total particle numbers $N_{\text{Na}}^{\text{tot}}{=}\sqrt{2 \pi}\sigma_{\text{Na}} N_{\text{Na}}^0 
/a$ and $N_{\text{NaLi}}^{\text{tot}}{=} \sqrt{2 \pi}\sigma_{\text{NaLi}} N_{\text{NaLi}}^0 
/a$ where the lattice constant, $a{=}\lambda/2$ and $\lambda {=} 1,596$ nm. As the averages weighted over a Gaussian, the effective particle numbers per pancake in Eq.~(\ref{eq:diffeqn1}) become
$N_{\text{NaLi}}^{\text{eff}}{=}N_{\text{NaLi}}^0 /\sqrt{2}$ and $N_{\text{Na}}^{\text{eff}}{=}N_{\text{Na}}^0 (\sigma_{ov}/ \sigma_{\text{NaLi}})$, where $1/\sigma_{\text{ov}}^2{=}1/\sigma_{\text{Na}}^2 {+} 1/\sigma_{\text{NaLi}}^2$. In Eq.~(\ref{eq:diffeqn2}), the effective particle numbers are $N_{\text{Na}}^{\text{eff}}{=}N_{\text{Na}}^0 /\sqrt{2}$ and $N_{\text{NaLi}}^{\text{eff}}{=}N_{\text{NaLi}}^0 (\sigma_{\text{ov}}/ \sigma_{\text{Na}})$.

We now discuss the volumes $V_{\text{eff}}$ and $V_{\text{ov}}$ in Eqs.~(\ref{eq:diffeqn1}) and (\ref{eq:diffeqn2}). Assuming a harmonic trap, one obtains $V_{\text{eff}} = \bar{\omega}_{\text{NaLi}}^{-3}(4\pi k_B T_{\text{NaLi}}/m_{\text{NaLi}})^{3/2}$ and $V_{\text{ov}} {=} (N_{\text{Na}}N_{\text{NaLi}})/(\int dV n_{\text{Na}}n_{\text{NaLi}}) {=}\bar{\omega}^{-3}_{\text{Na}}[(2\pi k_B T_{\text{Na}}/m_{\text{Na}}) \\ \times (1 {+}(\alpha_{\text{Na}}/\alpha_{\text{NaLi}})(T_{\text{NaLi}}/T_{\text{Na}}))]^{\frac{3}{2}}$ where the geometric mean of the sodium trap frequencies, $\bar{\omega}_{\text{Na}} {=} (\omega_x\omega_y\omega_z)^{1/3} \\ {=} 2\pi {\times} (325 {\cdot} 460 {\cdot} 45000)^{1/3}$ Hz and the polarizability ratio, $\alpha_{\text{NaLi}}/\alpha_{\text{Na}} = (m\omega^{2})_{\text{NaLi}}/(m\omega^{2})_{\text{Na}} = 2.8(4)$. 
 
However, the confinement in each pancake is strongly anharmonic . In our model, we use the actual beam geometry of a 1D lattice trap with a retro-reflected beam displaced by $2\; \text{mm}$ (which is ${\sim} 60 \%$ of the Rayleigh range) and  $43 \%$ power relative to the incoming beam. Both beams are linearly polarized and have the same beam waist of $43\; \mu\text{m}$. The trap depth is calculated from the measured trap frequencies. In the local density approximation, we obtain
\begin{equation}
   1/V_{\text{ov}} =\cfrac{\int dV e^{-\beta_{\text{Na}}U_{\text{Na}}(r, z)} e^{-\beta_{\text{NaLi}}U_{\text{NaLi}}(r, z)}}{[\int dV e^{-\beta_{\text{Na}}U_{\text{Na}}(r, z)}][\int dV e^{-\beta_{\text{NaLi}}U_{\text{NaLi}}(r, z)}]}
    \label{eq:overlap}\tag{S3}
\end{equation}
where $r$ is the radial coordinate, $z$ is the axial coordinate along the beam direction, $\beta_i = (k_B T_i)^{-1}$, and $U_i(r,z)$ is the potential of a single lattice site for the particle $i$. The integration limits in both coordinates are determined by the equipotential contour of the local maximum formed by gravity that tilts the trap in the y-direction.
We find that the harmonic overlap volume is smaller than the anharmonic volume by a factor of 1.8.

Using this model and calibration, we experimentally measured the absolute loss rate for the mixture in the non-stretched spin state to be $30 \%$ lower than the universal limit, which agrees with the prediction of universality within the experimental uncertainty of ${\sim}40 \%$.
    
The validity of the anharmonic model is experimentally cross-checked by observing cross-dimensional thermalization which occurs with a cross section well known via the  $s$-wave scattering length of sodium atoms \cite{na_scattering_length},
including the 2D correction of the elastic scattering rate \cite{dima_2d}.
For this, $V_{\text{eff}}$ is determined from the anharmonic trapping potential as
\begin{equation}
   1/V_{\text{eff}} = \cfrac{\int dV \text{e}^{-2\beta_{\text{Na}}U_{\text{Na}}(r, z)}}{[\int dV \text{e}^{-\beta_{\text{Na}}U_{\text{Na}}(r, z)}]^2}
    \label{eq:4}\tag{S4}
\end{equation}

We find that the observed thermalization rate differs from the model prediction by ${\sim} 10 \%$, which is within the uncertainty of the thermalization measurement, ${\sim}15 \%$.
    
When we use the same model for the trapping potential to obtain the two-body molecular loss rate coefficient, we find a factor of two discrepancy with the expected $p$-wave universal limit \cite{Note3}.
This is most likely due to a non-equilibrium distribution of molecules over the different ``pancakes'' due to the formation process.

In summary, in the main paper, we have eliminated the need for absolute values of the overlap density $N^{\text{eff}}_{\text{Na}}/V_{\text{ov}}$ by comparing loss rates around the Feshbach resonances to the loss rate of the mixture in the non-stretched spin state, which was predicted to be universal. We have validated this prediction by the calibration procedure and modeling described here. In other words, if we had not used the comparison with the non-stretched systems, we would have obtained the same results within the (now larger) uncertainties.

The loss rate coefficients reported here are somewhat different from previous values in Ref.~\cite{NaLiSympCool} where the anharmonicity in the trapping potential and the polarization purity of the absorption imaging light were not fully considered.

\subsection{Typical particle numbers $\& $ decay of sodium atoms}
\noindent  The typical particle numbers per pancake are 320 and 30 for Na and NaLi, respectively. With  the typical statistical uncertainty of the particle number, ${\sim} 13\%$, the decay of the sodium number is insignificant and we can treat the number as constant during the hold time.
    
However, near the peak of the strong resonance, the molecular signal depletes quickly while we wait for the field to fully settle after ramping it. Thus, we typically work in the regime where $N_{\text{Na}}/N_{\text{NaLi}} = 125/23\; {\approx} \;5$. In this case, we find decay curves of both sodium atoms and molecules fit well to the analytic solution of Eq.~(\ref{eq:diffeqn1}) and (\ref{eq:diffeqn2}) by ignoring the molecular two-body loss term which is two orders of magnitude smaller than the atom-molecule term. In this regime, the numerical solution of Eq.~(\ref{eq:diffeqn1}) and (\ref{eq:diffeqn2}) agrees well with the analytic solution: $N_{\text{Na}}(t) = -\text{D}/(e^{-\text{D}\Gamma t}/\text{C} - 1), N_{\text{NaLi}}(t) = \text{D}/(\text{C}e^{\text{D}\Gamma t} - 1)$, where $\Gamma$ is the loss rate and $\text{C} = N_{\text{Na}}(0)/N_{\text{NaLi}}(0), \text{D} = N_{\text{Na}}(0){-}N_{\text{NaLi}}(0)$. $N(0)$ is the initial particle number. When $D \approx N_{\text{Na}}(0)$, the sodium number is approximately constant, and we find the decay curve of molecules fits well to a simple exponential function, $N_{\text{NaLi}}(t)\approx N_{\text{NaLi}}(0)e^{-\Gamma t}$.

\subsection{Magnetic field inhomogeneity}
\noindent After a mixture of molecules and atoms is initially prepared at $745\; \text{G}$, the bias field is ramped to the target value in $15\; \text{ms}$. Due to eddy currents, an extra hold time of $15\; \text{ms}$ or more was added to allow the bias field to fully settle.  Due to magnetic field curvature, the sample experiences a magnetic field inhomogeneity of  $\lesssim 80\; \text{mG}$ near $1000\; \text{G}$. This upper bound is inferred from the narrowness of an RF spin-flip transition of sodium atoms.

\subsection{Temperature of Na and NaLi}
\noindent Inelastic losses occur predominantly at the highest densities and lead to increase in temperature, sometimes called anti-evaporation. Near the peak of the strong resonance, molecules heat up before we start acquiring a decay curve. We observe that within $\pm 2.5\; \text{G}$ around the peak, the molecule temperature is ${\sim} 1.9\; \mu\text{K}$ while the temperature of sodium atoms is at ${\sim} 1.6\; \mu\text{K}$, due to the large imbalance of particle numbers. This gives the temperature associated with relative radial motion, $T_{\text{rel}}{=}\mu(T_{\text{Na}}/m_{\text{Na}}+T_{\text{mol}}/m_{\text{mol}}){=}1.73(15)\; \mu\text{K}$ near the peak. Compared to the temperature away from the peak $T_{\text{rel}} {\sim} 1.55\; \mu\text{K}$, this is about $12 \%$ higher, and with the increased temperature, the fit result is still the same within the fit uncertainty. Further heating during the decay curve measurement is not observed. 

\begin{figure}
    \centering
	\includegraphics[width=86 mm, keepaspectratio]{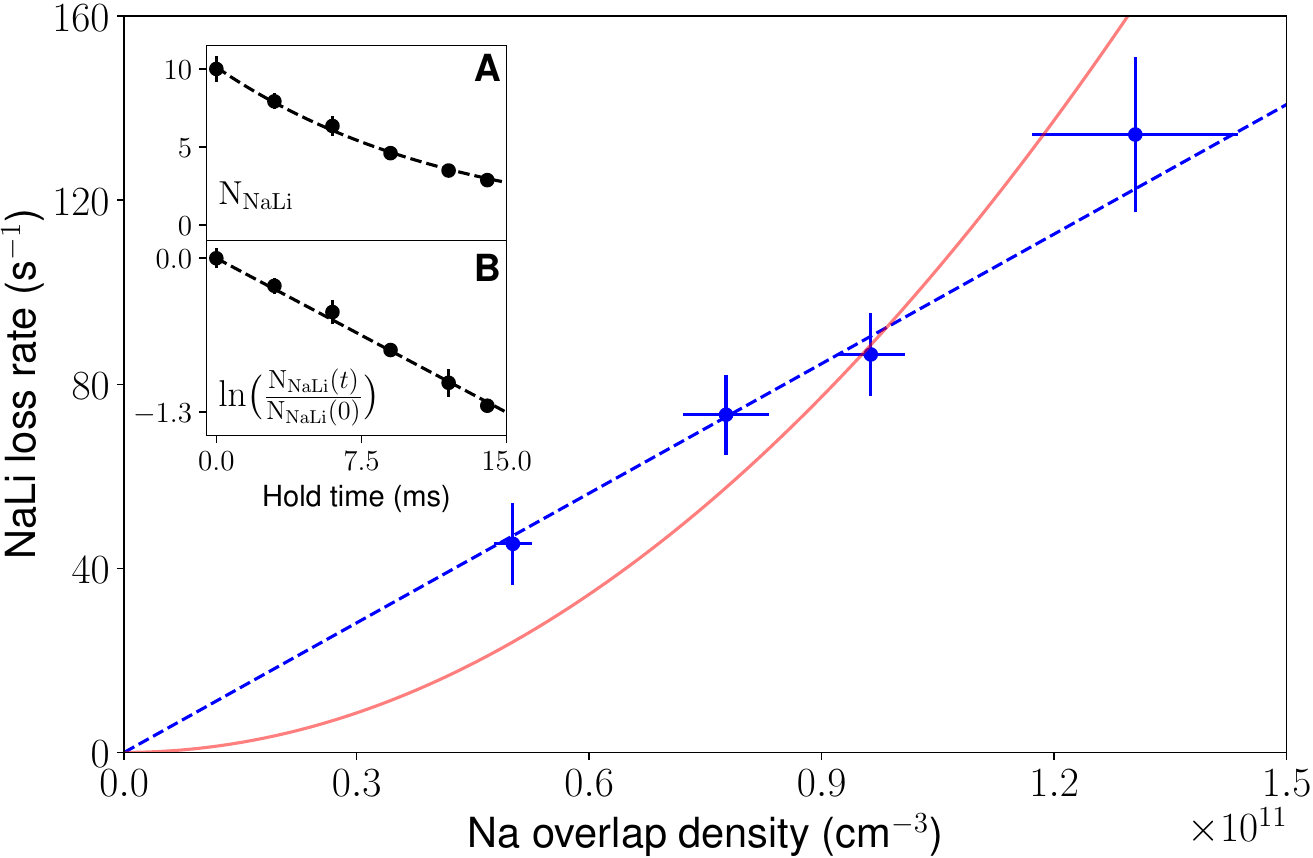}
	\caption{{\bf Verification of two-body loss mechanism.} The loss rate of molecules is shown as a function of the overlap density at $978.7\; \text{G}$, near the peak of the strong resonance and shows a linear dependence (blue dashed line). The red line is a quadratic fit.  The error bar of the y-data is one standard deviation. The error bar of the x-data is one standard error of the mean. In this figure, we are not including the calibration uncertainty for the Na density. The inset ($\textbf{A}$) shows a decay curve of molecules with overlap density $9.6(4) \times10^{10}\; \text{cm}^{-3}$, and ($\textbf{B}$) shows a linearized decay curve with a linearized exponential fit at the same overlap density.}
	\label{fig:twobody}
\end{figure}
\subsection{Verification of two-body loss mechanism}
\noindent We have treated the loss dynamics as purely two-body collisions. Because the scattering length near the peak gets very large, we checked that the decay is still dominated by two-body decay. Fig.~\ref{fig:twobody} shows indeed a linear dependency of the loss rate of molecules on the overlap density. It is acquired near the peak of the strong resonance, at $981\; \text{G}$ where Re$[\Tilde{a}] = -940\; a_0$. The blue dotted line is a linear fit ($\chi_{\text{red}}^2 = 0.3$) and the red line is a quadratic fit ($\chi_{\text{red}}^2 = 3.9$). This shows the observed losses have negligible contribution from three-body decay with $\text{n}^2_{\text{Na}}\text{n}_{\text{NaLi}}$ ($n_i$ is the density of particle $i$). The inset shows a linear fit of ln($\text{N}_{\text{NaLi}}(t))$ with $\chi_{\text{red}}^2 = 0.3$. In comparison, a linear fit of $\text{N}_{\text{NaLi}}(t)^{-1}$, which would reflect three-body decay with $\text{n}_{\text{Na}}\text{n}^2_{\text{NaLi}}$, gives $\chi_{\text{red}}^2 = 4.1$ (not shown in the figure).

\subsection{Mapping between Fabry-Perot model and threshold scattering formalism}
\noindent In the main text, we have shown that the results of the multichannel quantum defect model in the threshold regime can be expressed by the Fabry-Perot contrast. Here we show how the multiple reflections are treated in a S matrix formalism.
    
The quantum reflection and transmission of the scattering wavefunction in the long-range regime are formulated in terms of $r^{io} (r^{oi})$ and $t^{io} (t^{oi})$, the amplitude reflection and transmission coefficients for the wave travelling inward (outward) respectively, as introduced in \cite{gao_qdt_general, cong_qdt, PJ2}. 
In general, $r^{oi}$ and $r^{io}$ have different phases. The coefficient of the short-range reflection with loss is the short-range S-matrix, $S^c = \xi e^{i2\delta_s}$ where $\xi = (1-y)/(1+y)$ and $y$ is the quantum-defect parameter, as given in \cite{PJ2, cong_qdt_highpartial}. 
$\delta_s$ is the short-range phase shift with a finite range $0 {\leq} \delta_s {<} \pi$ \cite{PJ2}.
    
The transmission through the total potential is obtained as the coherent sum of multiple reflection pathways in the range  $R_{\text{short}} < R < R_{\text{long}}$ \cite{cong_qdt_highpartial}:
\begin{equation}
    \begin{split}
        T &= I|\sqrt{1 - \xi^2}t^{oi} (1 + r^{io}S^c + (r^{io}S^c)^2 + ...)|^2 \\
        &= I|\sqrt{1 - \xi^2}t^{oi}|^2 (1 - r^{io}S^c)^{-1}
    \end{split} 
    \label{eq:5}\tag{S5}
\end{equation}
where I is the influx.

Ref.~\cite{cong_qdt} gives an analytic solution of the phase factors for the reflection coefficients. For $s$-wave collisions with vdW interaction, the phases of $r^{oi}$ and $r^{io}$ are $\pi$ and $-\pi/4$ respectively. This simplifies the transmission as follows:
\begin{equation}
    T = I|t^{oi}|^2 \bigg(\frac{|\sqrt{1 - \xi^2}|^2}{|1 - |r^{io}|\xi e^{i(2(\delta_s - \pi/8))}|^2} \bigg) \equiv I|t_2|^2 C
    \label{eq:6}\tag{S6}
\end{equation}
This is identical to the transmission through two reflectors, given as Eq.~(1) in the main article, with $|t_2|=|\sqrt{1 - \xi^2}|$, $|r_1| = |r^{oi}|= |r^{io}|$, $\xi = |r_2|$. The negative round-trip phase $-\phi$ is twice the short-range phase shift $2\delta_s$ plus the phase shift due to the quantum reflection at long range $-\pi/4$ (i.e., $-\phi = 2\delta_s - \pi/4)$.

\subsection{Mapping of two-channel model to Breit-Wigner form}
\noindent Scattering resonances are described by a complex scattering length $\tilde{a}$ which moves across a circle in the complex plane as the resonance is crossed \cite{JH,JH_detail}. The most general resonance has six parameters (the two center coordinates and the radius of the circle, the position of the background on the circle, and the position/width of the resonance in magnetic field). Here we show that our two-channel model which has six parameters can be mapped to the standard Breit-Wigner form describing an isolated resonance.  In contrast, the one-channel Fabry-Perot model has only five parameters since the position and radius of the circle are constrained and depend only on $y$ and $\bar{a}$ \cite{PJ}.

The complex scattering length defined by the two-channel model with parameters $(B_{\text{res}}, \Delta, y, q, \Gamma_{\text{b}}, \bar{a})$ is given by:
\begin{equation}    
    \tilde{a}(B) = \bar{a} \bigg(s(B) + y \cfrac{1 + (1-s(B))^2}{i + y(1-s(B))}\bigg) \equiv \alpha(B) -i\beta(B)
	\label{eq:PJmodel}\tag{S7}
\end{equation}
The Feshbach resonance with a lossy bound state is described by:
\begin{equation}
    s(B) = q\bigg(1 - \cfrac{\Delta}{B-B_{\text{res}} -i\Gamma_{\text{b}}/2}\bigg)
    \label{eq:8}\tag{S8}
\end{equation}
We can re-write the two-channel model, Eq. (\ref{eq:PJmodel}), in the usual Breit-Wigner form \cite{JH} in terms of the complex background scattering length $\tilde{a}_{\text{bg}}$, the resonant scattering length, $\tilde{a}_{\text{res}}$, the resonance position, $B^{o}_{\text{res}}$, and the width, $\Gamma_{\text{inel}}$:
\begin{equation}
    \begin{split}
        \tilde{a}(B) &= \tilde{a}_{\text{bg}} + \cfrac{\tilde{a}_{\text{res}}}{2(B-B^{o}_{\text{res}})/\Gamma_{\text{inel}}- i} \\
        &\equiv \alpha(B) - i\beta(B)
    \end{split}
    \label{eq:JH_supp}\tag{S9}
\end{equation}
\indent The mapping between two parameter sets is as follows:
\begin{equation}
    \begin{split}
        &B^{o}_{\text{res}} = B_{\text{res}} +  \cfrac{qy^2\Delta(q-1)}{1 + y^2(q-1)^2}  \\
        &\Gamma_{\text{inel}} = \Gamma_{\text{b}} + \cfrac{2yq\Delta}{1 + y^2(q-1)^2} \equiv  \Gamma_{\text{b}} + \Gamma_{\text{c}}\\ \\ 
        &\tilde{a}_{\text{bg}} = \cfrac{y(2-q) + iq}{i + y(1-q)} \ \bar{a} \equiv \alpha_{\text{bg}} - i\beta_{\text{bg}} \\
        &\alpha_{\text{bg}} = \cfrac{q + (q-2)(q-1)y^2}{1 + (q-1)^2y^2}\ \bar{a} \\
        &\beta_{\text{bg}} = \cfrac{(2 + (q-2)q)y}{1 + (q-1)^2y^2}\ \bar{a} \\ \\
        &\tilde{a}_{\text{res}} = \cfrac{2}{\Gamma_{\text{inel}}} \cfrac{(1-y^2)q\Delta}{(i + y(1-q))^2}\ \bar{a} \equiv \alpha_{\text{res}} - i\beta_{\text{res}} \\
        &\alpha_{\text{res}} = \cfrac{-2q\Delta(1-y^2)(1-y^2(q-1)^2)}{(1 + (q-1)^2y^2)[(1 + (q-1)^2y^2)\Gamma_{\text{b}} + 2qy\Delta]}\ \bar{a} \\
        &\beta_{\text{res}} = \cfrac{4qy\Delta(q-1)(y^2-1)}{(1 + (q-1)^2y^2)[(1 + (q-1)^2y^2)\Gamma_{\text{b}} + 2qy\Delta]}\ \bar{a}
	\end{split}
    \label{eq:10}\tag{S10}
\end{equation}
where the width of the resonance $\Gamma_{\text{inel}}$ is the incoherent sum of two decay rates: the natural linewidth of the bound state, $\Gamma_{\text{b}}$ and the width determined by the coupling strength between the scattering state and the bound state, $\Gamma_{\text{c}}$. Note that both width factors have units of magnetic field.  The energy width is obtained by dividing by $\delta\mu$. Since the energy widths are positive definite, the signs of $\Gamma_{\text{b}}$ and $\Gamma_{\text{c}}$ are determined by the sign of $\delta\mu$. In our case, $\Gamma_{\text{b}}$ and $\Gamma_{\text{c}}$ are positive since $\delta\mu>0$. When $\delta\mu<0$, $\Gamma_{\text{b}}<0$ by definition, and $\Gamma_{\text{c}}<0$ since $q \cdot\Delta<0$ (see Sign constraints for parameters $q$ and $\Delta$).

The imaginary part of the scattering length is given by:
\begin{equation}
    \beta = \beta_{\text{bg}} + \cfrac{-\alpha_{\text{res}} (\Gamma_{\text{inel}}/2)^2+ \beta_{\text{res}}(\Gamma_{\text{inel}}/2) (B-B^{o}_{\text{res}})}{(B-B^{o}_{\text{res}})^2 + (\Gamma_{\text{inel}}/2)^2}
    \label{eq:11}\tag{S11}
\end{equation}
\indent It is straightforward to check that the peak height is $\beta = 1/y$ at $B {=} B_{\text{res}}$, as expected.

\indent For $y \ll 1$, $q \sim 1$, and $\Gamma_{\text{inel}} \approx \Gamma_{\text{c}}$:
\begin{equation}
    \beta(B) \approx \beta_{\text{bg}} + yq\Delta\bigg[\cfrac{q\Delta+ 2(1-q) (B-B^{o}_{\text{res}})}{(B-B^{o}_{\text{res}})^2 + (yq\Delta)^2}\bigg]\bar{a}
    \label{eq:12}\tag{S12}
\end{equation}
This shows that due to the second term in the numerator, $q>1$ and $q<1$ give different asymmetries in the lineshape. \\
\indent In contrast, for $y \ll 1$, $q \sim 1$, and $\Gamma_{\text{inel}} \approx \Gamma_{\text{b}}$:
\begin{equation}
    \beta(B) \approx \beta_{\text{bg}} + q\Delta \bigg[\cfrac{\Gamma_{\text{b}}/2 + 2y(1-q)(B-B^{o}_{\text{res}})}{(B-B^{o}_{\text{res}})^2 + (\Gamma_{\text{b}}/2)^2}\bigg]\bar{a}
    \label{eq:13}\tag{S13}
\end{equation}
In this case, the asymmetry is small and the lineshape can be well-approximated by a Lorentzian. 

\subsection{Sign constraints for parameters $q$ and $\Delta$}
\noindent By definition, $q$ and $\delta\mu \Delta $ have the same sign \cite{chengchin_feshbach} where $\delta\mu = \mu_{\text{Na${+}$NaLi}} - \mu_{\text{bound}}$ is the relative magnetic moment between the entrance channel and the closed channel bound state. In our case, $\mu_{\text{Na${+}$NaLi}} = 3\mu_{\text{B}}$ as the entrance channel is in the upper spin-stretched state where three electronic spins are aligned along the bias field direction. Thus, $\delta\mu = 4\mu_{\text{B}} (2\mu_{\text{B}})>0$, if the bound state we couple requires two(one) spins to flip. This constrains the signs of $\Delta$ and $q$ to be the same, i.e. $q\cdot \Delta {>}0$. The observed asymmetry of the loss lineshape requires $q{>}1$, which implies  $\Delta{>}0$.

\subsection{Thermal averaging}
\noindent At our temperature, $k_B T{<}\hbar\omega_{ax}$, the scattering is dominated by the axial motional ground-state channel \cite{dima_2d} and the saturation factor does not depend on temperature $f(k) = (1 + (\sqrt{\pi}/l_{o})^2|l|^2|\tilde{a}|^2+2(\sqrt{\pi}/l_o)|l|^2\beta)^{-1}$. However, the logarithmic correction factor,  $l {=}|(1 + (\tilde{a}/\sqrt{\pi}l_o)\text{ln}(B\hbar\omega_{ax}/\pi \epsilon))|^{-2}$ (where $\epsilon {=}\hbar^2q^2/2\mu$, $B{\approx}0.915$, and $l_o{=}\sqrt{\hbar/\mu\omega_{ax}}$) depends on the radial momentum $q$. In the main article, we used the mean thermal energy $k_BT$ for the radial relative kinetic energy $\epsilon$. A more accurate way is to perform a thermal average.

\indent The thermal average for the loss rate ratio $r(B) = f(k) l (\beta/ \bar{a})$ is given by: 
\begin{equation}    
    \braket{r(B)} = \cfrac{\beta f(k)}{\bar{a}}\cfrac{\lambda}{Z}\bigg(\int_{0}^{\infty} dq\; 2\pi q\; l(q) e^{\frac{-\hbar^2q^2}{2\mu k_B T}}\bigg)
    \label{eq:lossconst_thermal}\tag{S14}
\end{equation}
where $Z = \mu k_B T/\hbar^2$, and $\lambda = \sqrt{1 - e^{-2\hbar\omega_{\text{ax}}/k_B T}}$  is the normalization constant for the average over the axial direction \cite{PJ_2d}.

At our temperature $T{\sim}1.55\; \mu\text{K}$, the thermal average effect shifts the peak loss position by ${\sim}0.3\; \text{G}$, almost canceling the shift caused by the mean thermal energy, $\epsilon {=}k_B T$. Fit parameters except $B_{\text{res}}$ agree with the result without the thermal average within the fit uncertainties.

\begin{figure}
    \centering
	\includegraphics[width=86mm, keepaspectratio]{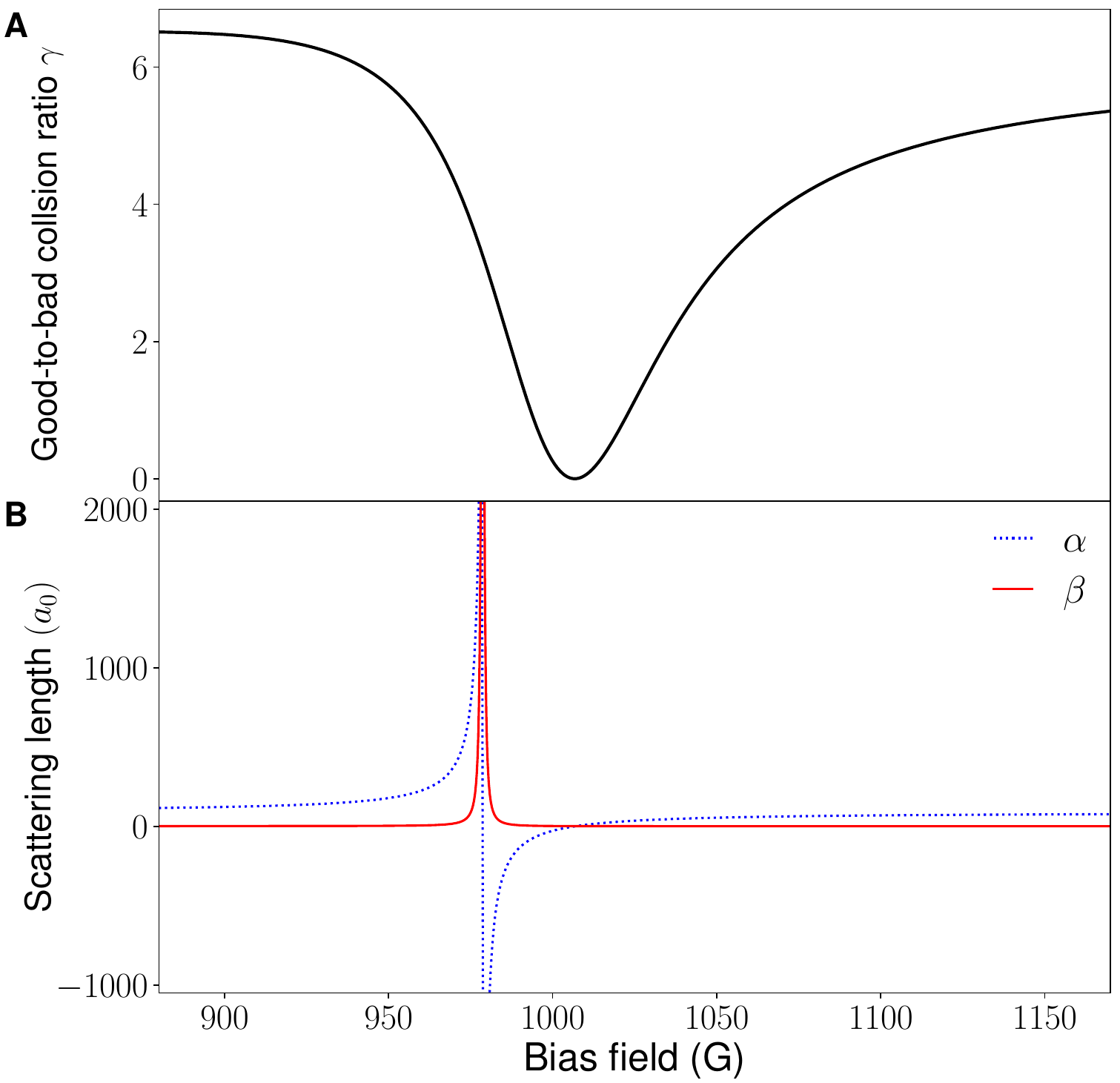}
	\caption{{\bf Ratio of elastic to inelastic collision rates near the strong resonance.} (\textbf{A}) The good-to-bad collision ratio, $\gamma{=}k(\alpha^2+\beta^2)/\beta$ as a function of the bias field, where $\alpha$ and $\beta$ are the real and and imaginary parts of the complex scattering length respectively, the relative wavevector $k=\mu v_{\text{rel}}/\hbar$ and $v_{\text{rel}} = \sqrt{8k_BT/\mu\pi}$ ($\mu$: the reduced mass between NaLi and Na, $T = 1.55 \mu$K).  (\textbf{B}) $\alpha$ and $\beta$ around the strong resonance at $978$ G. Note that the effect of the small resonance at $1030$ G has not been considered in the calculation of $\gamma$.}
	\label{fig:gamma}
\end{figure}

\subsection{Ratio of elastic to inelastic collision rates}
\noindent Figure \ref{fig:gamma} shows the good-to-bad collision ratio,
$\gamma{=}k(\alpha^2+\beta^2)/\beta$. It is maximized away from the Feshbach resonance, since near the resonance, the increase in the elastic scattering rate, which is proportional to ($\alpha^2+\beta^2$), is less than the increase in the inelastic scattering rate, which is proportional to  $\beta/k$. The minimum of $\gamma$ is not at the resonance
but at $1006.9$ G where  the real part of the scattering length, $\alpha$ crosses zero and the imaginary part, $\beta$ is close to it background value. Note that this plot uses only the parameters obtained from the analysis
of the resonance at $978$ G. In the previous work Ref.~$\textit{(30)}$, we demonstrated sympathetic cooling of NaLi molecules with Na atoms near $745$ G and measured a higher ratio of good-to-bad collisions of about $100$. This discrepancy is presumed to be caused by contributions from other Feshbach resonances and is the subject of future studies. Also, in Ref.~\cite{NaLiSympCool} we measured elastic collisions directly by observing thermalization, whereas here we inferred an elastic collision rate only from the analysis of observed inelastic rates.

\end{document}